\documentclass[fleqn,usenatbib]{mnras}

\usepackage{newtxtext,newtxmath}

\usepackage[T1]{fontenc}

\DeclareRobustCommand{\VAN}[3]{#2}
\let\VANthebibliography\thebibliography
\def\thebibliography{\DeclareRobustCommand{\VAN}[3]{##3}\VANthebibliography}

\usepackage{graphicx}	
\usepackage{amsmath}	
\usepackage{amsmath}
\usepackage[utf8]{inputenc}
\usepackage{caption}
\usepackage{booktabs}
\usepackage{multirow}
\usepackage{xcolor}
\usepackage{color}

\usepackage{enumerate}
\usepackage[utf8]{inputenc}
\usepackage{ wasysym }
\usepackage{graphicx}
\usepackage{bbm}
\usepackage{algorithm}
\usepackage{algpseudocode}
\usepackage{mathrsfs}

\title[Short title, max. 45 characters]{Searching for ring-like structures in the Cosmic Microwave Background}

\author[M. Lopez et al.]{
M. L\'{o}pez $^{1,2}$\thanks{E-mail: m.lopez@uu.nl}
P. Bonizzi$^{3}$
K. Driessens$^{3}$
G. Koekoek$^{4}$
J. A. de Vries$^{4}$
R. Westra$^{3,4}$
\\
$^1$ Institute for Gravitational and Subatomic Physics (GRASP), Department of Physics, Utrecht University, Princetonplein 1, 3584 CC Utrecht, The Netherlands. \\
$^{2}$ Nikhef, Science Park 105, 1098 XG Amsterdam, The Netherlands. \\ 
$^{3}$  Department of Advanced Computing Sciences, Maastricht University, Maastricht, the Netherlands. \\ 
$^{4}$ Department of Gravitational waves and Fundamental Physics, Maastricht University, Maastricht, the Netherlands.
}

\date{Accepted XXX. Received YYY; in original form ZZZ}

\pubyear{2022}

\begin{document}
\label{firstpage}
\pagerange{\pageref{firstpage}--\pageref{lastpage}}
\maketitle

\begin{abstract}

{In this research, we present an alternative methodology to search for ring-like structures in the sky with unusually large temperature gradients, namely Hawking points (HP), in the Cosmic Microwave Background (CMB), which are possible observational effects associated with Conformal Cyclic Cosmology (CCC). To assess the performance of our method, we constructed an artificial data set of HP, according to CCC, and we were able to retrieve $95 \%$ of ring-like anomalies from it. Furthermore, we scanned the \textit{Planck} CMB sky map and compared it to simulations according to $\Lambda CDM$, where we applied robust statistical tests to assess the existence of HP. Even though no significant ring-like structures were observed, we report the largest excess of HP candidates found at $\alpha = $1\% significance level for the analyzed sky maps (CMB at 70GHz, SEVEM, SMICA, and Commander-Ruler), and we stress the need to continue the theoretical and experimental research in this direction.}

\end{abstract}


\section{Introduction}

Numerous theories have been proposed and several experiments have been performed to answer one of the most fundamental questions of humanity: \textit{what are the dynamics of the very early and very late Universe?} The resulting answer involves the Standard Cosmological Model, $\Lambda CDM$, although there are some inconsistencies with other areas of Physics that some speculative theories try to solve (see discussion in \cite{RoadToReality|Penrose}).
In this research we explore the observational implications in the Cosmic Microwave Background (CMB) of a theory of Sir Roger Penrose named Conformal Cyclic Cosmology (CCC) \cite{CyclesOfTime|Penrose}, whose latest statistical analyses have been under discussion \cite{ApparentEvidenceForHP|Penrose, Re-evaluationgEvidenceforHP|Jow}. Moreover, we develop a method to assess its presence and potential location in the CMB.

The main premise of CCC is that the Universe is in a state of eternal inflation and its full evolution is divided in \textit{aeons} that are separated from each other by conformal transformations at the \textit{cross-over} \citep{CyclesOfTime|Penrose, RoadToReality|Penrose}.  In a remote future all matter will either be captured in supermassive black holes, that subsequently radiate into mass-less particles by \textit{Hawking evaporation}, or, as it is postulated by CCC, will experience a mass fade out to become mass-less with time (see \cite{ApparentEvidenceForHP|Penrose} for details). Eventually, this remote and mass-less universe will be conformally indistinguishable from the start of a new universe. Penrose argues that this resemblance is not coincidental and as such, the end of a universe can be mapped to the beginning of a new universe, making the history of the Universe a (possibly endless) cycle of \textit{aeons}. 

 Similarly to  $\Lambda CDM$, every \textit{aeon} described by CCC starts with a Big Bang and, due to the cosmological constant $\Lambda > 0$, ends with an exponential expansion in a remote future around $\sim 10^{100}$ years. 
The main differences between $\Lambda CDM$ and CCC are that Penrose does not consider a model with early Universe \textit{inflation} and that the entire history of the Universe is taken to be a succession of \textit{aeons}. In CCC, an \textit{aeon} is defined as the start of the Universe and the end of it. 

According to CCC, there are two types of events from the previous \textit{aeon} that could be observed in the present universe. One of these would be the effects of gravitational waves coming from inspiralling pairs of supermassive black holes and the other the effects of the evaporation of supermassive black holes, known as \textit{Hawking points}  (hereafter \textit{“HPs”}). In this research, we will focus our attention on the second type of event. 

At the end of an \textit{aeon}, mass-less particles are \textit{squashed conformally}  at the {cross-over}. Hence, all radiation from the evaporation of a black hole is concentrated into an {HP}, and it will travel into the subsequent \textit{aeon} ``heating'' the matter of the early Universe, leaving imprints in the Cosmic Microwave Background (CMB). Due to relativistic constraints and given the known expansion of the universe, the imprints are expected in the form of a \textit{Gaussian-like} distribution,  projected in the CMB as circular annuli that do not exceed an outer diameter of $\sim 4^{\circ}$, corresponding to a maximum outer radius of $\sim 0.035$ radians \cite{ApparentEvidenceForHP|Penrose}. {There is also an inner radius of the annuli since the supermassive black hole will have evaporated before the cross-over takes place. The size of this inner radius depends on how much time has passed between evaporation has been completed and the cross-over, and cannot be determined \textit{a priori}.}

To accept or reject CCC it is crucial to perform robust analyses
on the data from the CMB to verify the existence of {HPs}  and their locations in the sky. This study aims to search for anomalous annuli in the CMB temperature field data by comparing two different distributions: one population where {HPs} could potentially exist and another population where {HPs} are not present. The first population comes from the CMB sky map measured by \textit{Planck} (hereafter \textit{real CMB sky map}), while the second population is obtained via simulations according to the Standard Cosmological theory $\Lambda CDM$ (hereafter \textit{simulated CMB sky map}).  

This paper is organised as follows. A brief overview of related works is given in Section \ref{related_work}. In Section \ref{data} we describe the real data and the simulations of the CMB temperature field. Subsequently, we define our measurements (Section \ref{measurements}), we provide a pseudo-code of the algorithm to scan the CMB sky maps (Section \ref{algorithm}) and we describe the methodology employed to measure the quality of the simulations, the existence of HPs and their locations (Sections \ref{stats_c} and \ref{stats_a}, \ref{stats_b}). In Section \ref{results}  we report the results, and show the performance of our procedure.  Finally, in Section \ref{conclusion} we discuss our results and provide some conclusions.

\section{Related work: imprints of the previous aeon}\label{related_work}

Several studies aiming at testing CCC have been conducted since Penrose presented his theory. Most of these studies aimed to indirectly detect gravitational waves produced by the collision of supermassive black holes in the previous aeon. It is theorized that collisions and formations of supermassive black holes would happen in the same galaxies due to the abundance of matter. As a consequence of several mergers in the same region,   those remnants from the previous \textit{aeon} should be projected in the CMB as concentric rings with an angular diameter $\leq 40^{\circ}$. Although this type of event is not discussed in this study, we can extrapolate some search methodologies from it, and therefore we briefly describe them below.

In \cite{ConcentricCirclesINWMAP|Gurzadyan}, the authors explore Wilkinson Microwave Anisotropy Probe (WMAP) data to search for rings with low-temperature variance according to a certain threshold and large populations of annuli were encountered in some regions. 
The points selected in the CMB were centres of concentric rings and deviated $6 \sigma$  along the annuli from the Gaussian expectation of $\Lambda CDM$. In \cite{OnCCCPredictedConcentric|Gurzadya}, it was proposed the \textit{sky-twist} methodology to test whether the numerous centres of multiple low-variance rings depended upon them being circular rather than some other shape, as the existence of elliptical shapes in WMAP sky map would contradict CCC. However, it was observed that low-variance ellipses were nonexistent in the data. 

In \cite*{StructuresInMBR|Meissner} three different frequency bands with artificial Gaussian maps with the same harmonic spectrum were scanned and their results compared, while in \cite*{StructuresINPlanck|Meissner} the same study was carried out by using Planck data. The comparison was performed by means of a methodology proposed in \cite{ATailSensitiveTest|Meissner}, known as \textit{A-functions} (also presented in Section \ref{stats_a}).
{Due to a high computational cost only  100 statistical artificial maps were scanned and the authors found ring-like structures with $99.7\%$ confidence level}.

In \cite*{SearchingForConcentric|DeAbreu}, the analysis of \cite{ConcentricCirclesINWMAP|Gurzadyan} with WMAP and Planck data was repeated. The authors performed the same procedure not only on real maps, but also on sky maps sampled from Gaussian distributions containing the usual CMB anisotropy power spectrum, which is consistent with the predictions of $\Lambda CDM$, and the same results were obtained for both types of maps. Moreover, it is argued that the threshold chosen to select the low-variance rings in \cite{ConcentricCirclesINWMAP|Gurzadyan} was a particular search criterion and had no statistical significance. 

In \cite{ASearchForCOncentricCircles|Wehus} the same analysis as in \cite{StructuresInMBR|Meissner} was implemented, but instead of employing the {A-functions} to compare the real distribution against an artificial distribution, the authors developed a methodology based on matched filters and $\chi^2$ statistics to compare both distributions. Due to their results, the authors concluded that no imprints were present in the data. 

Regarding studies that aimed to detect {HPs}, in  \cite{ApparentEvidenceForHP|Penrose}, after scanning a  real CMB sky map and 10.000 simulated maps, the authors claimed to have found {HPs} in the real sky map at $99.98 \%$ confidence level according to the metrics of \cite{ATailSensitiveTest|Meissner}, in favour of CCC.  In \cite{Re-evaluationgEvidenceforHP|Jow}, the authors reported to have found {HPs} at $87 \%$ confidence level with the same metrics, concluding that this result is not significant enough. It is important to note that both studies employed large sky maps of  $N_{\text{side}}=1024$, where this parameter represents the resolution of the grid used according to Healpy \citep{healpix}.

\section{The Cosmic Microwave Background temperature field}\label{data}

The CMB offers us a look at the universe when it was about 1/36.000 of its present age. At that time, the Universe became transparent, emitting photons that have travelled freely ever since. However, not all photons that arrive at our antennas belong to the CMB. The main astrophysical sources of noise come from our Galaxy through different mechanisms: synchrotron radiation, free-free emission and spinning dust radiation, which dominates at low frequencies, and thermal dust, which dominates at higher frequencies. At $70$ GHz, the contribution of sources of noise is smaller \citep{ApparentEvidenceForHP|Penrose}, but there are still systematic errors in the CMB temperature determination that are not taken into account. For these reasons, many cleaning techniques had been developed to remove the foregrounds from the sky maps obtained from telescopes, such as Spectral Estimation Via Expectation-Maximization (SEVEM), Spectral Matching Independent Component Analysis (SMICA) and Commander - Ruler \citep{Planck:2013fzg}. 
In the present work, we employ the full missions maps cleaned from \textit{Planck} 2nd release\footnote{They can be downloaded from Infrared Science Archive on Planck Public Data Release 2 \url{https://irsa.ipac.caltech.edu/data/Planck/release_2/all-sky-maps/matrix_cmb.html}}, as well as the CMB sky map at $70$ GHz, and we refer to them as the \textit{real CMB sky maps}.

As we mentioned before, we aim to compare a population where HPs are potentially present, namely the real CMB sky maps, with a population where HPs are guaranteed to not exist, i.e. simulated CMB sky maps following $\Lambda CDM$, to later compute their similarities or dissimilarities.   Simulating the CMB temperature field is non-trivial, as this radiation encloses a large amount of information about the primordial Universe, which is strongly dependent on its shape and early composition. Fortunately, we can artificially obtain CMB sky maps with the Code for Anisotropies in the Microwave Background (CAMB) \citep*{CAMB|Howlett, CAMB|Lewis}. With CAMB we can compute the angular power spectrum of the simulation $C_{\ell_{\text{sim}}}$ according to the cosmological parameters obtained from \textit{Planck 2018} \citep{Planck2018VI}, and with \textit{Healpy} a temperature field is generated.  From the possible values that multi-pole $\ell$ could acquire, we exclude $\ell_0$ and $\ell_1$ and we only expand the CMB until $\ell=1500$, which is expected to have a negligible effect on the temperature, as proposed in \citep{ApparentEvidenceForHP|Penrose}. Due to the resolution of our maps being $N_{\text{side}}=1024$, to simulate Planck's camera resolution we smooth $C_{\ell_{\text{sim}}}$ with a Gaussian beam at $fwhm = 10 '$  according to  \cite{Planck2013XXIII} with \textit{Healpy} package \citep{Healpy|Zonca, Healpy|Gorski}. 

Because we are interested in testing the existence and the location of HPs in many simulated maps, we computed several artificial maps coming from the same angular power spectrum with \textit{Healpy}. The code that generates the simulated maps needs to distribute the temperature field following $\Lambda CDM$, with the assumption that the CMB sky map is almost a Gaussian distribution. As the distribution occurs with a random seed, no two maps are alike and we can simulate the artificial maps recursively. 
\section{The methodology}\label{methodology}

We propose a methodology to find ring-like anisotropies in the CMB. For this aim we compare two  distributions of measurements:  one where HPs could potentially exist (real CMB sky maps) and another one where they are not  present (simulated CMB sky maps in accordance to $\Lambda CDM$). To construct both distributions we need to scan all the different sky maps from Section \ref{data} and calculate specific metrics at each pre-defined sky location for annuli of different inner and outer radius $\{r_{\text{in}}, r_{\text{out}}\}$. In this study, we employed two metrics: normalized slope and Pearson correlation coefficient (both introduced in Section \ref{measurements}).

Once we have computed all values for a metric,  we perform  statistical tests to understand if temperature fluctuations are expected by $\Lambda CDM$ or are anomalies in accordance to CCC. For this, we measure the similarity or dissimilarity of both distributions of a metric, in the region where {HPs} are expected according to CCC, ($r_{\text{out}} \leq 0.035$ rad). Another crucial point is to assess the reliability of the simulations, i.e. we need to compare both distributions in the region where HPs are  not expected under the assumption that they are similar ($r_{\text{out}} \geq 0.035$ rad). Finally, if  HPs were to be found in the real CMB sky maps, we would locate them to compare them against the ones found in \cite{ApparentEvidenceForHP|Penrose}. 

This section is structured as follows: we provide a definition of the two metrics employed in \ref{measurements}, and we present a pseudo-code of the scanning algorithm in \ref{algorithm}. In \ref{stats_a} we give an overview of the statistical tests employed, and the underlying hypotheses. In Section \ref{stats_b} we present a methodology to compute the locations of {HPs} in the real CMB sky map,  if we were to find evidences of their existence. Finally, in Section \ref{stats_c} we assess performance of the algorithm with an artificial data set.

\subsection{Measuring the slope}\label{measurements}

According to CCC, {HPs} present a Gaussian-like distribution, but in the real CMB sky map they would be projected as 2-dimensional annuli, whose temperature gradient monotonically decreases as we move away from the center. {Around the region where the temperature gradient is maximum we can approximate $T(d)$ to depend linearly on the angular distance $d$, in the local neighborhood of the maximum gradient of $T(d)$.} With this idea in mind, several authors searched for {negative slopes} of the temperature around given sky directions (see \cite{ApparentEvidenceForHP|Penrose} and \cite{Re-evaluationgEvidenceforHP|Jow} for details). {HPs} should show the most negative normalized slopes of the population,  but it is not known \textit{how much negative}.

To find such slopes, we can employ common least squares, where we define $a$ as the slope of the linear relationship and $b$ as the intercept of the line $T = ad +b$. Because we intend to scan the CMB sky maps to search for annuli of different sizes $\{r_{\text{in}}, r_{\text{out}}\}$, one needs to keep in mind that a change in size implies a change in the number of pixels
and hence different variances for the slope. To take this into account, in \cite{ApparentEvidenceForHP|Penrose} the authors define the measurement needed to search for {HPs} as the normalized slope, i.e. $\hat{a}=a/\sigma_a$, where $\sigma_a$ is {the standard deviation of the distribution of measured slopes of all annuli of the same size present in the sky map. In this research we also use the Pearson correlation coefficient $r$ {between the angular distance $d$ and the temperature $T$} as a measurement to search for {HPs}, because the whole measurement of a {HP} depends on the assumption {that these two variables} are linearly correlated.}

\subsection{Scanning the CMB}\label{algorithm}

{We scanned 1,004 sky maps of $\approx 12.5$ million pixels each to search for HP. For this aim, we built a Python code employing \textit{Healpy} and \textit{Numba} packages for efficient computing \citep*{Healpy|Gorski,Healpy|Zonca,Numba|Lam}, and we provide its pseudo-code in  Algorithm \ref{alg:algo_1}.}

\begin{algorithm}
\caption{Computation of measurement $m$}\label{alg:algo_1}
\begin{algorithmic}[1]
\Require{Temperature $T$,  $c$ centers, annulus size $\{r_{\text{in}}, r_{\text{out}}\}$}.
\State{Initialise $M=\emptyset$}
\ForAll {$c$}
\ForAll {$\{r_{\text{in}}, r_{\text{out}}\}$}
\State {Get pixels:}
\State {\hspace{3mm}$\mathcal{P}(c, r_{\text{in}}, r_{\text{out}}) = \text{masked sky map}\cap \mathcal{A}(c, r_{\text{in}}, r_{\text{out}})$ }

\If {$\# \left( \mathcal{A}(c, r_{\text{in}}, r_{\text{out}}) \setminus \textit{masked} \right) > 75$} 
\State {Get temperatures  $T(\mathcal{P}(c, r_{\text{in}}, r_{\text{out}}))$} 
\State {Calculate  angular distances $d(\mathcal{P}(c, r_{\text{in}}, r_{\text{out}}))$ } 
\State {Calculate the measurement $m(c, r_{\text{in}}, r_{\text{out}})$} 
\State {$M = M \cup \{m(c, r_{\text{in}}, r_{\text{out}}\}$}  
\EndIf
\EndFor
\EndFor
\State { \hspace{-1mm}\textbf{return} $M  $}
\end{algorithmic}
\end{algorithm}

The task of Algorithm \ref{alg:algo_1} is to  scan a {sky map} for each pre-defined annulus, defined as 
\begin{equation}
    \small{\mathcal{A}(c, r_{\text{in}}, r_{\text{out}}) \equiv \{ x \notin \mathcal{X} \text{ }|\text{ } r_{\text{in}}< |x-c| < r_{\text{out}}\}},
\end{equation}
{for a given center $c =\{l,\text{ } b\}${, where $l$ and $b$ represent the galactic longitude and latitude respectively,} and a certain inner and outer radius $\{r_{\text{in}}, r_{\text{out}}\}$. Defining the width $\epsilon = r_{out} - r_{\text{in}}$, the range of annuli explored in this research is $\epsilon \in [0.005, 0.03]$ with a step of $0.005$ and $r_{\text{in}} \in [0.000, 0.040]$ with a step of $0.025$, which is half of the  $r_{\text{in}}$ step used in  \cite{ApparentEvidenceForHP|Penrose}. The motivation behind the choice of these parameter ranges is to have a similar number of sizes where HP are both expected and not expected to be found theoretically.}

\begin{table}
\caption{{Number of pre-defined locations ($C$) explored for each mask.}}
\label{tab:total_locs}
\begin{tabular}{cccc}
\hline
\textbf{}       & $\min(C)$ & $\max(C)$ & $\mu(C)$ \\ \hline
SEVEM           & 3,182  & 10,383 & 7,062   \\
SMICA           & 2,832 & 28,7151 & 13,781  \\
Commander-Ruler & 3,645  & 23,357 & 13,332  \\
Galactic belt   & 12,320 & 12,640 & 12,487  \\ \hline
\end{tabular}
\end{table}

{To avoid undesired contribution of the Galactic center we used a Galactic belt mask defined as $\mathcal{X} = \{0<l< 2 \pi ,-\frac{\pi}{4} <b<\frac{\pi}{4}\}$, as proposed in \cite{ApparentEvidenceForHP|Penrose}. Furthermore, we also employed pre-defined masks $\mathcal{X}$ of SEVEM, SMICA and Commander-Ruler, which are more restrictive and cover the undesired contribution from point sources outside of the Milky Way galaxy. }

{Because different masks $\mathcal{X}$ cover different amount of pixels, we describe in Table \ref{tab:total_locs} the number of pre-defined centers $c$ where we compute measurements, which we define as $C$. Due to the large number of sizes explored, we present the minimum, maximum, and mean ($\mu$) of $C$. }

At each location we obtain the pixels of annuli with different sizes $\{r_{\text{in}}, r_{\text{out}}\}$ as

\begin{equation}
    \small{\mathcal{P}(c, r_{\text{in}}, r_{\text{out}})= \textit{masked sky map }\cap \mathcal{A}(c, r_{\text{in}}, r_{\text{out}})}.
\end{equation}

Computationally, we obtain the pixels of the inner circle $\mathcal{P}_{in}$, and the pixels of the outer circle $\mathcal{P}_{out}$ to get the inverse intersection    {$\mathcal{P}(c, r_{\text{in}}, r_{\text{out}}) =(\mathcal{P}_{in}(c, r_{\text{in}}) \text{ } \cap \text{ }  \mathcal{P}_{out}(c, r_{\text{out}}))'$}.
{We obtain the temperature of the pixels $T(\mathcal{P}(c, r_{\text{in}}, r_{\text{out}}))$ and the angular distance  $d(\mathcal{P}(c, r_{\text{in}}, r_{\text{out}}))$ from the pixels to the center $c$.\footnote{Instead of using the built in function  \textit{healpy.rotator.angdist()} to compute the angular distance $d(\mathcal{P})$, we implement our own function with \textit{Numba} package, which is twice as fast.} Finally, we calculate 
the values of the slope ${a}$ and the Pearson coefficient $r$ for each $(c, r_{\text{in}}, r_{\text{out}})$. }

 {We define a family of annuli $\mathscr{A}(r_{\text{in}}, r_{\text{out}})$ to be the set of all annuli with the same size but different centers $c$,}

\begin{equation}
    \small{\mathscr{A}(r_{\text{in}}, r_{\text{out}}) \equiv \{\mathcal{A}(c, r_{\text{in}}, r_{\text{out}})\text{ } \text{ }|\text{ }    \forall \text{ } c  \}},
\end{equation}

{Furthermore, we define $\sigma_a= \sigma({\mathscr{A}(r_{\text{in}}, r_{\text{out}})})$, as the standard deviation of the slopes of all the annuli that belong to the same family.  Thus, we normalize the slope $a$ as $\hat{a}=a/\sigma_a$  (see Section \ref{measurements} for details).}

\subsection{Statistical methods for the absence of HPs and the reliability of the simulations}\label{stats_a}

In this subsection we present different  hypothesis  (labeled in Roman numerals) and two statistical methodologies (labeled in arabic numbers) to search for HPs in the CMB. Once we have calculated the values of the normalized slope $\hat{a}$ and the Pearson coefficient $r$ for each $(c, r_{\text{in}}, r_{\text{out}})$, and for all the real and simulated CMB sky  maps, we would like to see what are the similarities between them. Therefore, given a {sky map}, we define $\mathscr{M}(r_{\text{in}}, r_{\text{out}})$ to be the set of all the measurements  for a fixed size $\{r_{\text{in}}, r_{\text{out}}\}$ as,

\begin{equation}
    \small{\mathscr{M}_{r_{\text{in}}, r_{\text{out}}}^{\text{sky map}} =\{m(c, r_{\text{in}}, r_{\text{out}}) \text{ } |\text{ }    c \in {\rm I\!R}^2 \text{ }\}}
\end{equation}

and we test the similarity of the measurements from the real ($ \mathscr{M}^{real}_{r_{\text{in}}, r_{\text{out}}}$) and the simulated sky maps ($ \mathscr{M}^{sim}_{r_{\text{in}}, r_{\text{out}}}$) in two different cases:

\begin{enumerate}[I]
\item \textit{Testing the reliability of simulated CMB sky maps for $r_{\text{out}} \geq 0.035:$ $\mathscr{M}^{real}_{r_{\text{in}}, r_{\text{out}}} \sim \mathscr{M}^{sim}_{r_{\text{in}}, r_{\text{out}}}$ on average.}
    
    For this case and according to CCC, the ring-like structures associated with HPs should not occur in the real CMB sky maps, so a similarity is expected between $\mathscr{M}^{real}_{r_{\text{in}}, r_{\text{out}}}$ and $\mathscr{M}^{sim}_{r_{\text{in}}, r_{\text{out}}}$, on average, with a p-value $p \geq 0.01$. Results of this analysis are reported in Section \ref{quality}. 
    
    \item  \textit{ Testing the absence of HPs for $r_{\text{out}} \leq 0.035$ $\mathscr{M}^{real}_{r_{\text{in}}, r_{\text{out}}} \geq \mathscr{M}^{sim}_{r_{\text{in}}, r_{\text{out}}}$ on average.}
    
    For this regime and according to CCC, HPs should be observable through their associated ring-like structures in the real CMB sky maps. Setting the non-existence of HPs as our null-hypothesis, we expect to find no proof of a larger amount of extreme negative slope measurements in $\mathscr{M}^{real}_{r_{\text{in}}, r_{\text{out}}}$, or that $\mathscr{M}^{real}_{r_{\text{in}}, r_{\text{out}}} \geq \mathscr{M}^{sim}_{r_{\text{in}}, r_{\text{out}}}$ on average with a p-value $p \geq 0.01$.  Inability to reject this hypothesis would  imply that manifestations of HPs are not observed in the CMB or that they are expected by $\Lambda CDM$. On the contrary, rejecting this hypothesis ($p < 0.01$) could give an indication towards the existence of HPs. Results of this analysis are reported in Section \ref{absence}.
\end{enumerate}

We calculated the statistics for both cases with two different methods, with the aim of comparing their performance.

\begin{enumerate}
    \item {\textit{Kolmogorov-Smirnov  (KS test):}} {
 
    is a non-parametric test of equality of probability distributions.  We employ the two-sample version of the test which relies on the \textit{empirical cumulative distribution functions} (eCDFs) \citep*{dekking2005modern}. Assuming that two independent samples $X_1, \dots, X_m$ and $Y_1, \dots, Y_n$ are drawn with eCDF's $F_m(t)$ and $G_n(t)$, the KS test finds the maximum distance between both {sampled probability distributions}.
    As explained in \cite{Pratt}, the test statistics are given by}
    \begin{equation}
        D_{mn} = \max{|G_n(t) - F_m(t)|}
        \label{eq:equal}
    \end{equation}
    \begin{equation}
        D_{mn}^{+} = \max{(G_n(t) - F_m(t))}
        \label{eq:bigger}
    \end{equation}
    \begin{equation}
        D_{mn}^{-} = \max{(F_m(t) - G_n(t))}
        \label{eq:smaller}
    \end{equation}
    
    Eq. \ref{eq:equal} is called the two-sided statistic, since it measures if $G = F$. Eq. \ref{eq:bigger} and \ref{eq:smaller} are called one-sided statistics, since they test if $G \geq F$ and $G \leq F$, respectively.
    Some advantages of KS test are that it does not depend on the underlying cumulative distribution function being tested and it is an exact test, with several desirable properties as consistency, and good power in detecting a shift in the median of the distribution \citep{Probability_and_Stats|Pearson}, \cite{Janssen}.  In our particular framework, 
    it is important to note that when testing cases I and II the statistics used is defined according to Eq. \ref{eq:equal} and \ref{eq:bigger}, respectively.
    
\end{enumerate}

When performing the KS-test for a  fixed annulus size $\{r_{\text{in}}, r_{\text{out}}\}$, we computed the similarity between a single real map $\mathscr{M}^{real}_{r_{\text{in}}, r_{\text{out}}}$ against $N=10^3$ simulated maps $\mathscr{M}^{sim}_{r_{\text{in}}, r_{\text{out}}}$, and we obtained a set of p-values. We define the set of p-values as, 

\begin{equation}
    \small{P_{\text{i}}(r_{\text{in}}, r_{\text{out}})=\{ p (x, i) \text{ }|\text{ } x \in \mathscr{M}^{sim}_{r_{\text{in}}, r_{\text{out}}}\} \text{ for } i \in \{1,4\}},
\end{equation}

{where $i$ refers to the four real {sky maps} presented in Section \ref{data}.
From a single set $P_{\text{i}}(r_{\text{in}}, r_{\text{out}})$ we calculated the confidence interval (CI) $\mu(P_{\text{i}}(r_{\text{in}}, r_{\text{out}})) \text{ } \pm \text{ } \varepsilon(P_{\text{i}}(r_{\text{in}}, r_{\text{out}}))$  at a confidence level of {$99.99 \%$}. If for a given size $\{r_{\text{in}}, r_{\text{out}}\}$ we were to find $\mu(P_{\text{i}}(r_{\text{in}}, r_{\text{out}})) < 0.01$ this could give an indication of the existence of HPs. }
\\
\\
The other method that has been commonly used in the search of HPs in the real CMB sky map \citep{ApparentEvidenceForHP|Penrose, Re-evaluationgEvidenceforHP|Jow} is based on proportion of simulated sky maps generating data as "exceptional" as the real sky map:

\begin{enumerate}
    \setcounter{enumi}{1}

    \item \textit{A-functions:} in \cite{ATailSensitiveTest|Meissner}, the author claimed that the main disadvantage of KS is that it tends to be more sensitive to the bulk of the distribution than to the tails, and the A-functions were proposed to overcome this problem. These functions are defined as follows,
\begin{equation}\label{eq:plus}
    \small{A^+(r_{\text{in}}, r_{\text{out}}) = -\frac{j}{N} \sum_{i=1}^{N}log(1-F_{\hat{a}}(r_{\text{in}}, r_{\text{out}})^j)},
\end{equation}
\begin{equation}\label{eq:minus}
    \small{A^-(r_{\text{in}}, r_{\text{out}}) = -\frac{j}{N} \sum_{i=1}^{N}log(1-[1-F_{\hat{a}}(r_{\text{in}}, r_{\text{out}})]^j)},
\end{equation}
where $F_{\hat{a}}(r_{\text{in}}, r_{\text{out}})$ is the cumulative distribution function (CDF) of the normalized slope $\hat{a}$ for all centers $c$, and $j$ is a positive real number that represents the relative weight of tails of the distribution. Distributions with an excess of extreme positive or negative values should give a large $A^+({r_{\text{in}}, r_{\text{out}}})$ or $A^-({r_{\text{in}}, r_{\text{out}}})$, respectively. 
\end{enumerate}

To compare the performance of the A-functions with the KS test, we followed they same approach as in \cite{ApparentEvidenceForHP|Penrose} and \cite{Re-evaluationgEvidenceforHP|Jow}. The methodology is as follows: the {A functions} are computed setting $j=10^4$ for each $\mathscr{M}^{sim}_{r_{\text{in}}, r_{\text{out}}}$ and $\mathscr{M}^{real}_{r_{\text{in}}, r_{\text{out}}}$, returning $\{A^+_{\text{sim}}(r_{\text{in}}, r_{\text{out}}),A^-_{\text{sim}}(r_{\text{in}}, r_{\text{out}})\}$ and $\{A^+_{\text{real}}(r_{\text{in}}, r_{\text{out}}),A^-_{\text{real}}(r_{\text{in}}, r_{\text{out}})\}$. 

\begin{itemize}
\renewcommand{\labelitemi}{\scriptsize \Circle}
    \item Whenever $A^-_{\text{sim}}(r_{\text{in}}, r_{\text{out}}) > A^-_{\text{real}}(r_{\text{in}}, r_{\text{out}})$, the counter $N^-(r_{\text{in}}, r_{\text{out}})$, representing the number of simulated maps with an excess of extreme negative values, is increased by $1$.
    \item When $A^+_{\text{sim}}(r_{\text{in}}, r_{\text{out}}) > A^+_{\text{real}}(r_{\text{in}}, r_{\text{out}})$, the counter $N^+(r_{\text{in}}, r_{\text{out}})$ for simulated extreme positive values is increased.
\end{itemize}

Distributions with an excess of extreme values should give a large $A^+(r_{\text{in}}, r_{\text{out}})$ and $A^-(r_{\text{in}}, r_{\text{out}})$. Therefore, if for a certain size we found a counter $N^+(r_{\text{in}}, r_{\text{out}}) \sim 0$, this would mean a large excess of positive  measurements, which is not expected by CCC. If  $N^{-}(r_{\text{in}}, r_{out }\leq 0.035) \sim 0$, it reflects that $\mathscr{M}^{real}_{r_{\text{in}}, r_{\text{out}}}$ has the largest population of negative extreme measurements, which could imply the existence of HPs.

\subsection{Statistical methods for the location of HPs}\label{stats_b}

The {A-functions} and the KS both test the presence or absence of HPs, but not their locations. 
To locate HPs we propose another analysis, whose aim is to store the location information of each anomaly.

\begin{enumerate}[I]
  \setcounter{enumi}{2}
  \item \textit{Location of HP candidates (HPC) at $\alpha$ significance level}
  
  Because we have simulated $N=10^3$ CMB sky maps, for a given center $c$ and a size $\{r_{\text{in}}, r_{\text{out}}\}$ we have a set of $N$ simulated measurements $m(c, r_{\text{in}}, r_{\text{out}})$ (either normalized slopes or  Pearson correlation coefficients), which we can define as:

\begin{equation}
\begin{aligned}
\mathbbmtt{M}(c,r_{\text{in}}, r_{\text{out}}) ={} & \{m_i(c, r_{\text{in}}, r_{\text{out}})\text{ }| \\
      & i \in 
      \text{ simulated sky maps} \}
\end{aligned}
\end{equation}

{We sort this set by increasing value and we define a certain significance level $\alpha$, that indicates the amount of extreme values that we expect in the tails. In the left tail of the simulated distribution $\mathbbmtt{M}(c,r_{\text{in}}, r_{\text{out}})$ we expect $\alpha \times N/2$ values. Thus, we take $t(c, r_{in}, r_{out}, \alpha)$, which is defined as the $\alpha \times N/2 th$ value of our sorted set,  to be the threshold for the detection of $HPs$.}

When scanning the $ith$ real CMB sky map for a certain  location $c$ and size $\{r_{\text{in}}, r_{\text{out}}\}$, if the  measurement $m_i(c, r_{\text{in}}, r_{\text{out}})$ {is smaller {(more negative)} than $t(c, r_{in}, r_{out}, \alpha)$} we consider this annulus $\mathcal{A}(c, r_{\text{in}}, r_{\text{out}})$ a HPC. Otherwise, it will not be considerate a candidate (NHPC). Note that the annulus is a function of $\{r_{\text{in}}, r_{\text{out}}\}$, so we are able to locate it in the sky map. Finally, we compute the rate of HPC, defined as the number of HPC over the total number of locations $c$ explored in percentage. We expect to find a HPC rate of $\alpha /2$ due to statistical fluctuations, {but larger rates would imply the existence of HPs}.
\end{enumerate}

A similar procedure can be employed to measure the population in the upper tail, just by ordering $\mathbbmtt{M}(c,r_{\text{in}}, r_{\text{out}})$ by decreasing value. It is important to note that this procedure does not only  point at HPC if they are anomalies in the data, but also locates them in the real CMB sky map.

\subsection{Statistical methods for the reliability of the methodology}\label{stats_c}

\begin{figure}

{
\includegraphics[width=0.9\columnwidth]{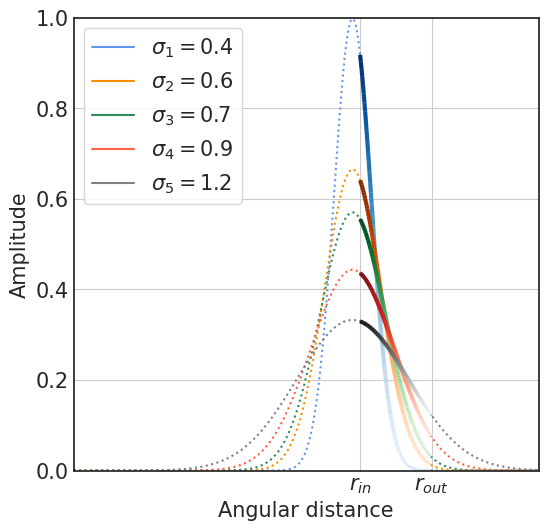}
}\hfill

{
\includegraphics[width=1\columnwidth]{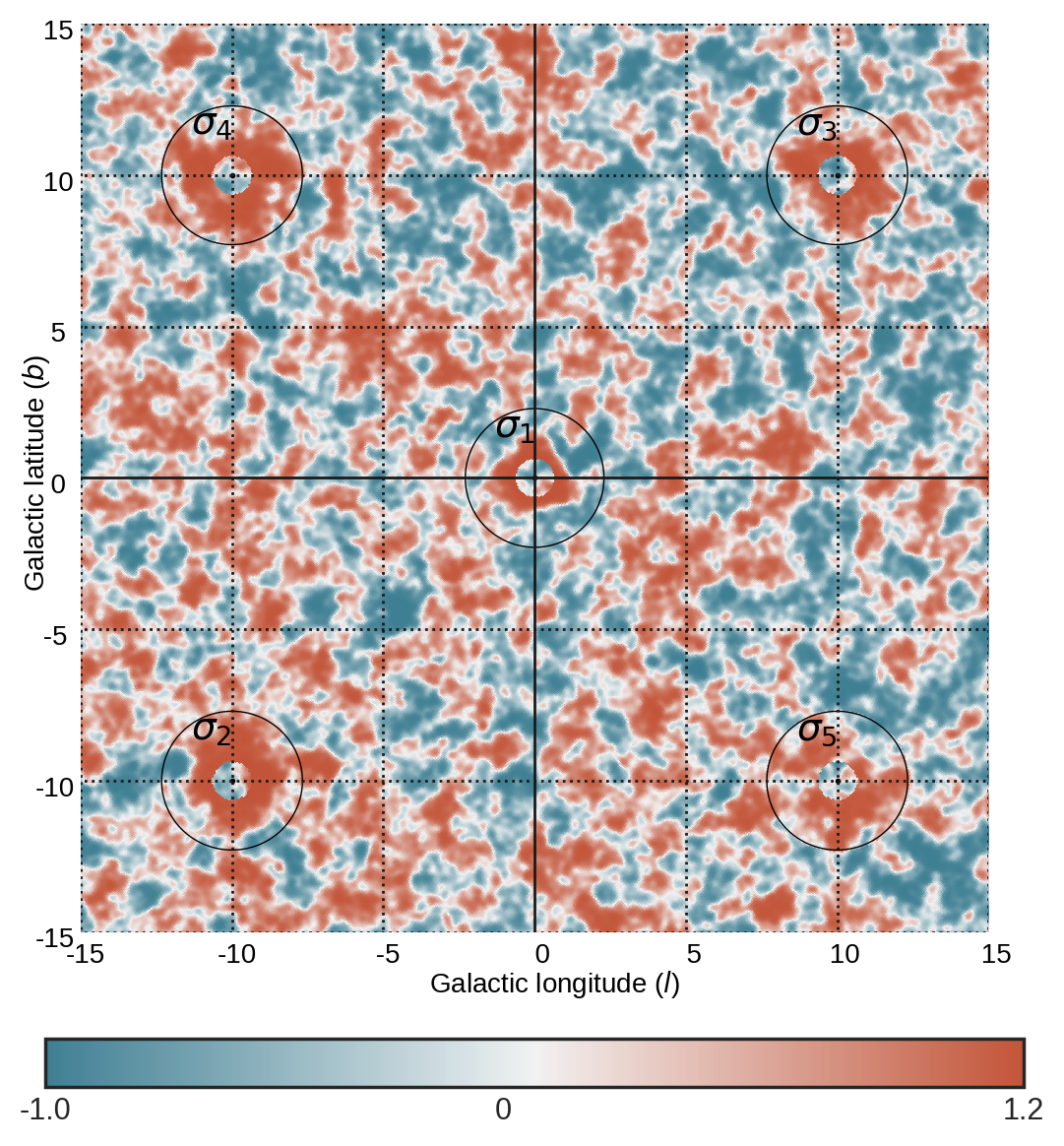}
}\hfill

\caption{{(Top) Gaussian distributions employed to construct AA. The solid line represents the portion of the Gaussian that has been rotated in the vertical axis to form a surface of revolution. Afterwards, this 3D surface was projected in a 2D simulated sky map. (Bottom) {Example of AA projected in galactic Cartesian coordinates $(l, b)$ measured in degrees that represent the angular distance. Each AA is of size $\{r_{in} = 0.0075, r_{out} = 0.0225\}$ and it is labeled according to the width of its generative portions in the top pannel.}} }
\label{fig:AR_gaussian}
\end{figure}

As a further proof of the performance of our methodology, we want to test whether we are able to capture ring-like structures with an artificial data set by employing the same methodology introduced in Section \ref{stats_b}. 
{To address this, we generated $1000$ simulated CMB sky maps, and we re-scaled them\footnote{We employ the rescaling function \texttt{preprocessing.MinMaxScaler()} from \texttt{scikit learn} \citep{scikit-learn}. Note that the rescaling function is applied to each individual sky map before injecting the AA.} in range $[-1, 1]$. Afterwards, we inject $110$ artificial annuli (\textit{AA}) per sky map.} Because we are unaware of how visible {HPs} are, {we created 5 different types of {AA} with different amplitudes to assess performance. The amplitude of each type of AA is inversely proportional to its width $\sigma$ and its radius $r$ has a finite size $\in [r_{in}, r_{out}]$, as it has been illustrated in Fig. \ref{fig:AR_gaussian} (top)}. After injecting AA, the upper-bound of the temperature field increases due to the excess that we have introduced, as it is reflected in the colorbar of Fig. \ref{fig:AR_gaussian} (bottom).

Note that re-scaling the sky maps individually would not affect the detection of AA, since our algorithm searches for strong linear correlations in the input data. Moreover, one must note that HP are Gaussian-like structures, so not strictly Gaussian. However, to construct a mock data set we approximated their slope to a Gaussian function. 

In Fig. \ref{fig:AR_gaussian} (top) we plot in solid line the portions of the Gaussians employed as a function of the number of pixels, to create different types of {AA}. {These portions are employed to generate Gaussian radial profiles, which are projected in the 2-dimensional simulated sky maps and injected at random sky positions. As an example, in Fig. \ref{fig:AR_gaussian} (bottom) {we represent in galactic Cartesian coordinates $(l, b)$ different AA of a fixed size labeled according to the portion employed (measured in degrees;} see top pannel for details).} {Note that the size of {AA} is sampled at random from the uniform distribution of sizes defined in Section \ref{algorithm}, with a limit $r_{\text{out}} \leq 0.07$ rad. } Furthermore, the upper-bound of the temperature field is $\ge 1.0$ because of the contribution of AA.

A single simulation is cloned $5$ times to insert a single type of {AA} per sky map. We stored their locations and scanned the maps, obtaining a population of measurements of normalized slope $\hat{a}$ and Pearson coefficient $r$.  After that, we built a confidence interval of the population at a certain $\alpha$ significance level to select the most negative measurements. 
Because each extremely negative measurement has a certain location $c$ and size $\{r_{\text{in}}, r_{\text{out}}\}$, we can count how many of these corresponding annuli intersect a single AA. If a single extremely negative annulus intersects an AA, then the AA is detected and labeled as such. Otherwise, we classify it as a \textit{fake AA}.     

We measured performance of our method in terms of the confusion matrix \citep{confusion_matrix}.  It is important to note that the amount of false positive (FP) is determined by the significance level $\alpha$.  {Since we are only interested in correctly classifying true AA (i.e., in true positive), we measured performance in terms of True Positive Rate (TPR), False Negative Rate (FNR), Positive Predictive Value (PPV) and False Discovery Rate (FDR) (see \cite{confusion_matrix} for details).}
In particular, for each CMB sky map we only compute TPR and PPV, because FNR and FDR can be derived from them, respectively.

\section{Results}\label{results}

{In section \ref{synthetic} we show the performance of our method when we employ a synthetic data set. 
In section \ref{quality} we present the results of testing the reliability of the simulations, while in section \ref{absence} we test the absence of HPs in the CMB sky map. For these last two sections, we employ the real sky maps SEVEM, SMICA, and Commander-Ruler, covered with their pre-defined masks, and CMB at 70 GHz sky map, masked by the Galactic belt mask, as well as $10^3$ simulated CMB maps. Note that when comparing the simulations against a certain real map, all-sky maps are covered with the same mask.}

\subsection{On the generation and detection of artificial HPs}\label{synthetic}

\begin{figure}
    \centering
    \includegraphics[width=0.45\textwidth]{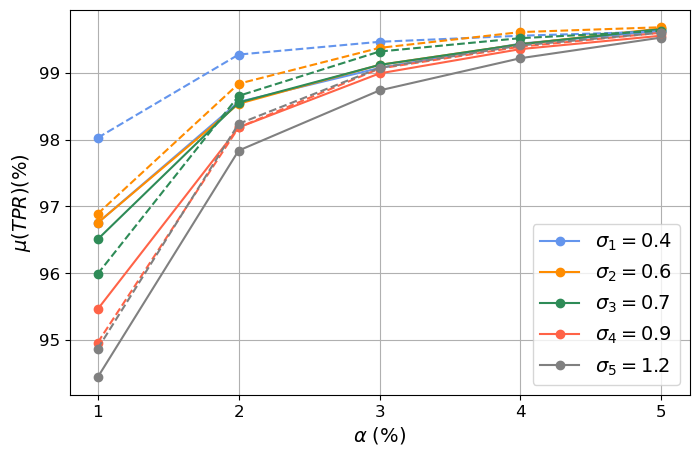}
    \caption{Average TPR as a function of the significance level $\alpha$ for different Gaussian rings. The solid lines represent the results of Pearson coefficient $r$ and the dotted lines are the results of the normalized slope $\hat{a}$.}
    \label{fig:TPR}
\end{figure}

\begin{figure}
    \centering
    \includegraphics[width=0.45\textwidth]{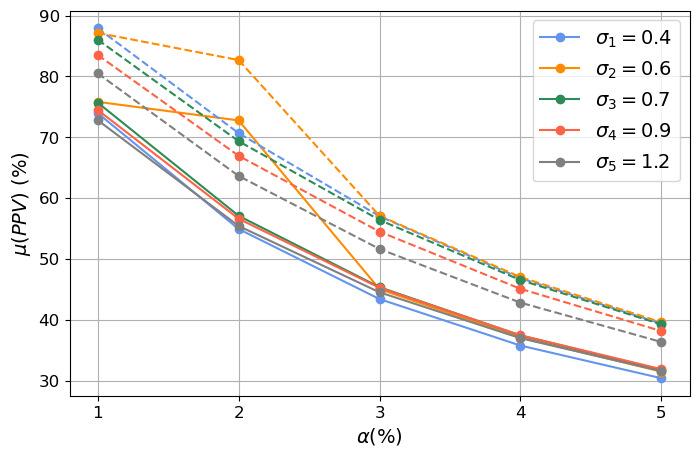}
    \caption{Average PPV as a function of the significance level $\alpha$ for different Gaussian rings. The solid lines represent the results of Pearson coefficient $r$ and the dotted lines are the results of the normalized slope $\hat{a}$.}
    \label{fig:PPV}
\end{figure}

As explained in Section \ref{stats_c}, we simulated $10^3$ artificial maps.  Once we generated the maps, we scanned them with our procedure and we computed for each map the normalized slope and Pearson correlation coefficient (see \ref{stats_c}). {In Fig. \ref{fig:TPR} and Fig. \ref{fig:PPV} we plot the mean TPR and the mean PPV as functions of the significance level $\alpha$ for {the different widths $\sigma$ (see Fig. 1 top pannel})}.

From Fig. \ref{fig:TPR} can observe that as we increase $\sigma$, the TPR and the PPV decrease. We speculate that for large $\sigma$, the TPR of the normalized slope $\hat{a}$ would be slightly better than the one of the Pearson coefficient $r$, because the Pearson coefficient points out the strongest correlations in the data. Note that, FNR has a complementary behaviour to TPR. In Fig. \ref{fig:PPV}, we conclude that the Pearson coefficient $r$ is less precise than the normalized slope $\hat{a}$, with a $\sim 20 \%$ difference for all $\alpha$. Nevertheless, with both measurements, we can detect most of the rings (around $95 \%$ TPR). 

\subsection{On the reliability of the simulated CMB maps}\label{quality}

To test the quality of the simulations we compared $10^3$ $\mathscr{M}_{r_{\text{in}}, r_{\text{out}}}^{\text{sim}}$ against  $\mathscr{M}_{r_{\text{in}}, r_{\text{out}}}^{\text{real}}$  of a selected real sky map in the region where  $r_{\text{out}} \geq 0.035$, i.e. where HP are not expected to be present according to CCC. For this aim, we use the approach introduced in Sec \ref{stats_a} with a significance level $\alpha=0.01$. After performing the test we obtain a distribution of $10^3$ p-values, and we compute its CI at $99.99\%$ confidence level. To have a reference mean p-value, we repeated the same procedure on simulated maps only, by comparing a simulated sky map against $10^3$ simulations. 
Assuming that the simulations of CMB are very similar to the real sky maps, it would be expected that for this test we would obtain $\mu(P_{i}) \geq 0.01$ for a given $ith$ sky map, meaning that the simulated maps are not significantly different from real CMB sky maps.

\begin{figure*}
  \includegraphics[width=\textwidth]{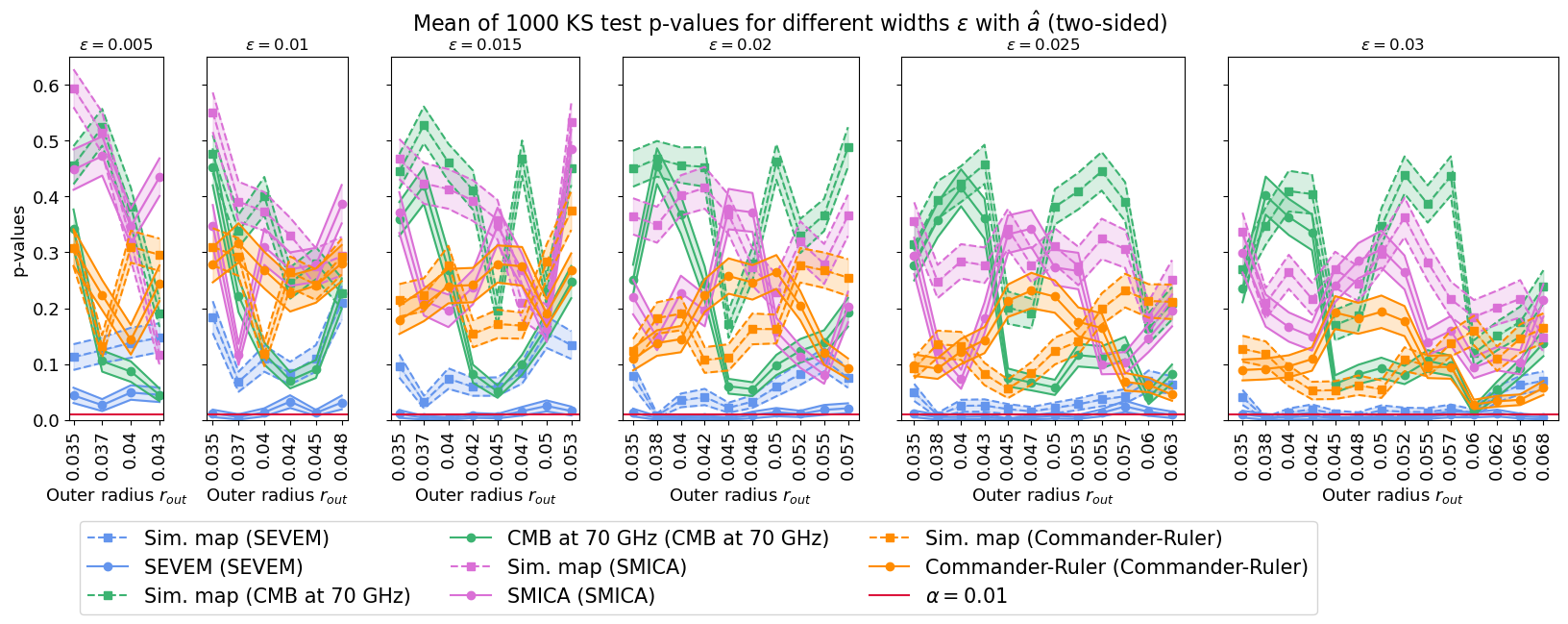}
  \caption{{Mean p-values at 4 standard deviations as a function of the outer radius where HP is not expected theoretically. {For this computation $10^3$ simulated sky maps were compared against a single simulated sky map according to $\Lambda CDM$ (dashed lines) and {a real sky map} (solid lines).} {All sky maps were masked according to the pre-defined mask obtained for their analysis,  except CMB at 70 GHz which was masked with the Galactic belt mask, as reflected in parenthesis in the legend.}}}\label{fig:quality_norm}
\end{figure*}

In Fig. \ref{fig:quality_norm} we represent the CI of the resulting p-values as a function of the outer radius of the annuli explored for the comparisons of real-simulated (solid lines) and simulated-simulated (dashed lines) sky maps. {Note that each pair of real-simulated and simulated-simulated sky maps have been masked with the same pre-defined mask.}
{From the results reported in Fig. \ref{fig:quality_norm}, it can be noticed that in general, we cannot reject the hypothesis that the pairs real-simulated and simulated-simulated come from the same distribution, except for SEVEM sky maps. Another exception, but marginal, is the real CMB at 70GHZ sky map, where for the size $\{r_{in}, r_{out} \} = \{0.057, 0.060\}$ in particular (far most right panel of Fig. \ref{fig:quality_norm}), we obtain a p-value $0.1674 \pm 0.0070$, whose lower bound is lower than the predefined threshold. One may think that this may be due to the low number of pixels in the annuli. However, in this region, we explored large annuli of $\sim 13,000$ pixels. It is relevant to note that Commander-Ruler might seem to have similar behaviour as CMB at 70GHZ at the same size,  but its lower bound does not intersect $\alpha = 0.01$.}

{The results of SEVEM seem to indicate that there is a strong influence of the mask employed. It would be interesting to further investigate the influence of this mask in the context of anomalous searches in the CMB sky map.}

\subsection{On the absence of HPs in the CMB}\label{absence}

To test the absence of HPs in the CMB, we compared again $10^3$ samples $\mathscr{M}_{r_{\text{in}}, r_{\text{out}}}^{\text{sim}}$ against $\mathscr{M}_{r_{\text{in}}, r_{\text{out}}}^{\text{real}}$ of a selected real sky map with KS test, where we used the approach II explained in Sec \ref{stats_a} with a significance level $\alpha = 0.01$.

Overall, we did not find any $\mu(P_{i}) <0.01$, which suggests that HP are not present in the CMB. However, a possible explanation for this result might be that there is not enough HP to produce a significant shift with respect to the reference distribution. 

For completeness, we calculated the counters $N^-_{r_{\text{in}}, r_{\text{out}}} \text{ and } N^+_{r_{\text{in}}, r_{\text{out}}}$ for the normalized slope $\hat{a}$, with the {A functions} setting $j=10^4$. As we discussed before, this methodology was proposed to give more importance to the tails of the distribution, because KS test gives more weight to the bulk instead (see Section \ref{stats_a}). {When computing  $N^-_{r_{\text{in}}, r_{\text{out}}} \text{ and } N^+_{r_{\text{in}}, r_{\text{out}}}$ we did not find statistically significant results for $\alpha = 1 \%$, i.e. we did not find an annulus size $\{r_{\text{in}}, r_{\text{out}}\}$ for which at most 10 simulated sky-maps outperformed the real sky-maps. Thus, we were unable to reproduce the results from \cite{ApparentEvidenceForHP|Penrose}, in which the authors reported to have identified HPs at a $99.98 \%$ confidence level.}

\begin{figure*}

  \includegraphics[width=\textwidth]{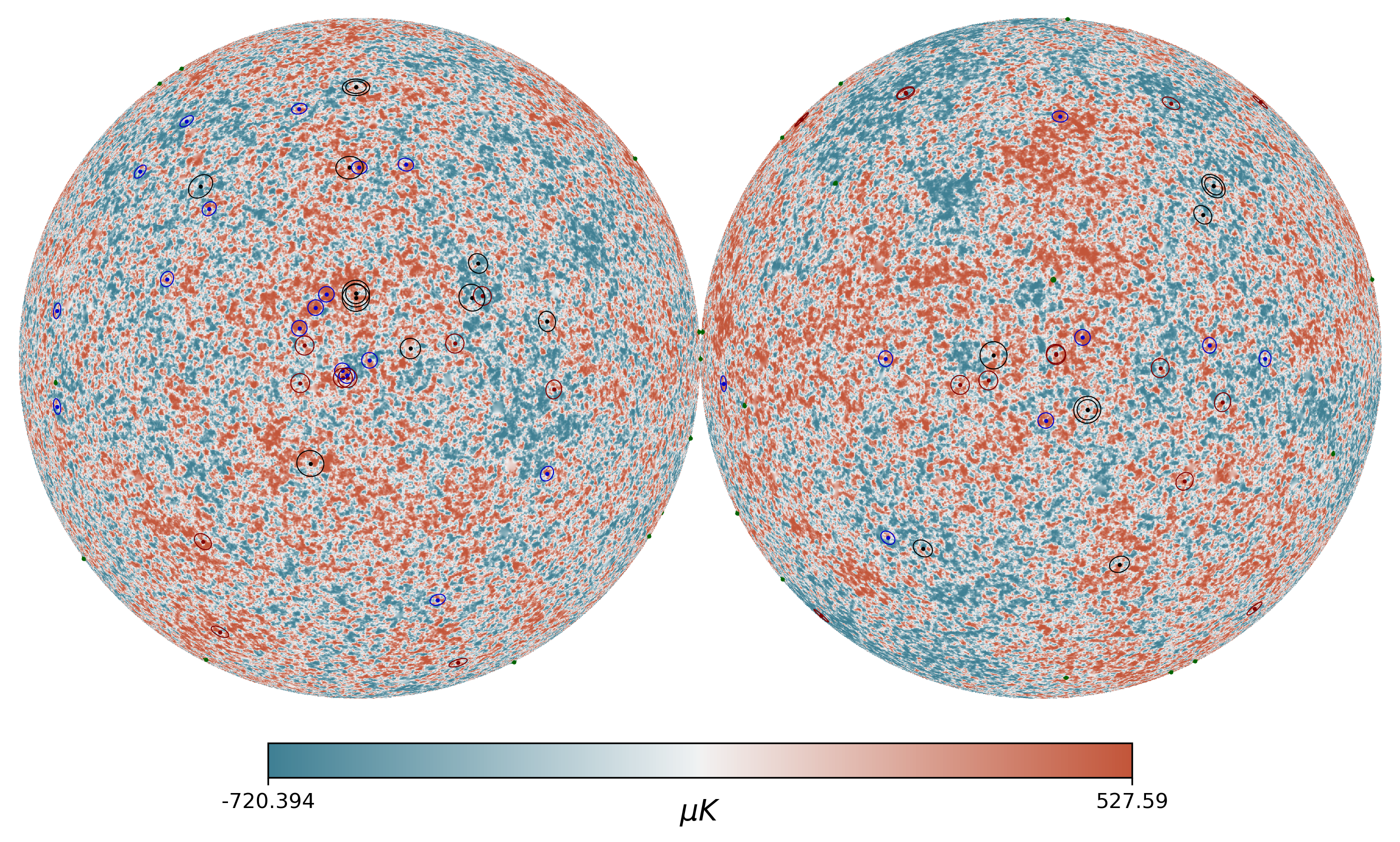}
  \caption{{HPC obtained with normalized slope $\hat{a}$ measurement at $1\%$ significance level, represented in orthographic projection: in black we plotted the most negative annuli presented in \citep{ApparentEvidenceForHP|Penrose}}, while in dark blue, dark green and dark red the HPC found by our confidence interval in CMB at 70 GHZ, SMICA and Commander-Ruler, respectively. The background CMB sky map corresponds to SMICA sky map, for illustration.}\label{fig:orthonormal}
\end{figure*}
\subsection{On the location of HPs in the CMB}

Finally, we built a confidence interval to obtain the most negative measurements, as we described in Section \ref{stats_b}, following the assumption that the simulations represent the real CMB sky maps. In  \cite{ApparentEvidenceForHP|Penrose} and \cite{CyclesOfTime|Penrose} it is stated that HPs should have extremely negative measurements \footnote{We also explored the extremely positive measurements for completeness but no interesting results were found.}, although no theoretical threshold is available for this value. {Therefore,  we computed different confidence intervals for $\alpha =[1 \%, 5 \%]$ for CMB at 70 GHz, SMICA and Commander-Ruler sky maps, since they are similar to the simulations (see Section \ref{quality} for details).} 

{Without considering these results as proof for HPs, we report on {the largest rates of HPC found} at $\alpha=1 \%$ for these sky maps. For the normalized slope $\hat{a}$ one expects a {HPC rate} of $0.5 \%$. The maximum {rates} obtained were of $0.734 \%$ for ($r_{\text{in}}=0.018$, $ r_{\text{out}}=0.023$) in CMB at 70GHz, $0.781 \%$ for ($r_{\text{in}}=0.003$, $ r_{\text{out}}=0.007$) in SMICA, and $0.903 \%$ for ($r_{\text{in}}=0.018$, $ r_{\text{out}}=0.028$) in Commander-Ruler.  We do not consider these deviations as exceptional and
believe that reporting on the significance of these deviations would be over-interpreting the data.}

{Fig. \ref{fig:orthonormal} shows in black the most negative annuli indicated by the authors in \cite{ApparentEvidenceForHP|Penrose}, and in dark blue, dark green and dark red 25 of the most extreme HPC found by our methodology at $\alpha = 1\%$ for CMB at 70GHz, SMICA and Commander-Ruler, respectively. As it can be observed, the annuli of \cite{ApparentEvidenceForHP|Penrose} and ours do not coincide,  due to the different methodologies employed. Notoriously, some locations of \cite{ApparentEvidenceForHP|Penrose}, CMB at 70GHz and Commander-Ruler partially overlap and others are in a close neighbourhood, despite employing different masks. It would be interesting to study these locations in detail in the future, as well as the influence of the masks used for this type of analysis.}
\section{Conclusion}\label{conclusion}

The main goal of this work was to search for ring-like anomalies in the CMB. Because we are searching for linear relationships between the measured temperature $ T$ and the angular distance $d$ in an annulus, we measured the normalized slopes $\hat{a}$ and Pearson coefficients $r$ for different $\{r_\text{in}, r_\text{out}\}$ sizes. The measurements were performed on real and simulated CMB sky maps with different masks. We compared both distributions of measurements to draw conclusions about the reliability of the simulations and the absence of HPs in the CMB. Furthermore, to measure the performance of our method with a mock data set, we constructed an artificial data set of ring-like structures and we detected artificial annuli with a $TPR \sim 95 \%$. 
 
{In this study, when measuring the quality of the simulations employed, we observed a mismatch between the real CMB maps and the simulations, in the region where ring-like structures associated with HPs are not expected when we employed the SEVEM sky map. This behaviour was not observed when we used CMB with Galactic belt mask, SMICA or Commander-ruler's masks.}

We also postulated the absence of the ring-like manifestations of HPs in the CMB, and we were unable to statistically reject this hypothesis by employing the Kolmogorov-Smirnov test, and thus we could not reproduce the results of \cite{ApparentEvidenceForHP|Penrose}. We think that these negative results may be because, if HPs exist, they may be rare in the real CMB map.

{To treat HPs as anomalies in the CMB, from simulated sky maps we generated CI for the normalized slope and Pearson correlation coefficient at different significance levels $\alpha$. When setting the confidence interval at $\alpha \%$ significance level we were expecting (in absence of HPs) HPC rates of at most $\alpha/2$ due to the randomness of the data. {We searched therefore for HPC rates larger than this value and found them for the normalized slope $\hat{a}$ at $ \alpha = 1 \%$ significance level in the CMB at 70 $GHz$, SMICA and Commander-Ruler sky map, which implies that there are more negative measurements in the real CMB sky maps than in the simulations.} Moreover, we located the most extreme values as can be observed in Fig. \ref{fig:orthonormal}.}

{Finally, we foresee a possibility, albeit speculative, to apply the methodology presented in this article to search for imprints of HPs in the Cosmological Gravitational-wave Background (CGB) of unresolved binary black hole (BBH) inspirals. The HP’s imprint on the energy distribution of the very early universe should have had a similar influence on the distribution of primordial black holes as it did on the formation of ring-like distributions in the CMB. The details of the formation and merger rates of primordial BBH is subject of ongoing theoretical investigation. However, it will be interesting to see whether such rings found in the CGB share the same centres as the candidates found in the CMB. If so, it would constitute additional evidence that the rings found in the CMB are not statistical artefacts, and would serve as an independent indication for the existence of HPs. The theoretical and observational feasibility of this suggestion, should be further investigated.}

\vskip 2mm
\section*{Acknowledgements}
We would like to express our great appreciation to Sir Roger Penrose for proposing this line of research. We gratefully thank Q. Meijer for the fruitful and inspiring discussions during this study, and S. Caudill for the critical reading of the manuscript and her
constructive inputs. This project was supported by Nikhef Laboratory, and we would like to thank the Nikhef computing group. M.L.  is supported by the research program  of  the Netherlands Organisation for Scientific Research (NWO).

\section*{Data Availability }

The data underlying this article are available in Planck Legacy Archive (PL) \citep{dataset}, at \url{https://www.cosmos.esa.int/web/planck/pla}. The datasets were obtained from observations with Planck (\url{http://www.esa.int/Planck}), an ESA science mission with in- struments and contributions directly funded by ESA Member States, NASA, and Canada.

\bibliographystyle{agsm}
\bibliography{references}


\bsp	
\label{lastpage}
\end{document}